\documentclass[12pt,preprint]{aastex}
\usepackage{multirow}
\usepackage{booktabs}
\usepackage{graphicx}
\usepackage{amsmath}
\usepackage{longtable}
\usepackage{rotating}
\usepackage{pdflscape}
\usepackage{color}

\newcommand{\msun}{M_\odot}
\newcommand{\mj}{M_{\rm J}}
\newcommand{\rp}{r_{\rm p}}
\newcommand{\mplanet}{M_{\rm p}}

\begin{document}
\title{Observational Signatures of Planets in Protoplanetary Disks II: Spiral Arms Observed in Scattered Light Imaging Can be Induced by Planets}

\shorttitle{Density Waves Excited by Planets}

\shortauthors{Dong et al.}

\author{Ruobing Dong\altaffilmark{1,2,4}, Zhaohuan Zhu\altaffilmark{3,4}, Roman R. Rafikov\altaffilmark{3}, James M. Stone\altaffilmark{3}}

\altaffiltext{1}{Lawrence Berkeley National Lab, Berkeley, CA 94720, rdong2013@berkeley.edu}
\altaffiltext{2}{Department of Astronomy, University of California at Berkeley, Berkeley, CA 94720}
\altaffiltext{3}{Department of Astrophysical Sciences, Princeton University, Princeton, NJ 08544}
\altaffiltext{4}{Hubble Fellow}

\clearpage

\begin{abstract}

Using 3D global hydro simulations coupled with radiative transfer calculations, we study the appearance of density waves induced by giant planets in direct imaging observations at near infrared wavelengths. We find that a $6\mj$ planet in a typical disk around a $1M_\odot$ star can produce prominent and detectable spiral arms both interior and exterior to its orbit. The inner arms have (1) two well separated arms in roughly $m=2$ symmetry, (2) exhibit $\sim10-15^\circ$ pitch angles, (3) $\sim180-270^\circ$ extension in the azimuthal direction, and (4) $\sim150\%$ surface brightness enhancement, all broadly consistent with observed spiral arms in the SAO~206462 and MWC~758 systems. The outer arms cannot explain observations as they are too tightly wound given typical disk scale height. We confirm previous results that the outer density waves excited by a $1\mj$ planet exhibit low contrast in the IR and are practically not detectable. We also find that 3D effects of the waves are important. Compared to isothermal models, density waves in adiabatic disks exhibit weaker contrast in surface density but stronger contrast in scattered light images, due to a more pronounced vertical structure in the former caused by shock heating and maybe hydraulic jump effect. To drive observed pairs of arms with an external companion on a circular orbit, a massive planet, possibly a brown dwarf, is needed at around [$r\sim0.7\arcsec$, PA$\sim10^\circ$] (position angle PA from north to east) in SAO~206462 and [$r\sim0.6\arcsec$, PA$\sim10^\circ$] in MWC~758. Their existence may be confirmed by direct imaging planet searches.

\end{abstract}



\section{Introduction}\label{sec:intro}

Detecting forming planets in protoplanetary disks is crucial for constraining the theory of planet formation, as it directly addresses two key questions: when and where do planets form. However, it is difficult to directly detect signals from forming planets in protoplanetary disks, and methods for indirect detection are necessitated.

In protoplanetary disks, planets can excite spiral density waves through gravitational disk-planet interactions \citep{kley12}. Density waves have much larger physical scale and can be much more prominent in observations than planets themselves. Once detected, they can be the smoking gun of the embedded unseen planets. Recently, high angular resolution direct imaging observations at near-infrared (NIR) wavelengths have found spiral-arm-like features in protoplanetary disks around young Herbig AeBe stars SAO~206462 \citep{muto12, garufi13} and MWC~758 \citep{grady13,benisty15}. These observations obtained polarized intensity (PI=$\sqrt{Q^2+U^2}$, the linear polarization component in the scattered light) images as a way to suppress the central starlight \citep{hinkley09,quanz11,hashimoto11}. These disks are relatively face-on, and show a pair of grand-design spiral arms often extending over $180^\circ$ in the azimuthal direction. These arms are located at several tens to a couple of hundred AU from the center, and they are nearly in $m=2$ rotational symmetry. At this moment resolved mm observations of these objects suffer from insufficient spatial resolution, and could not pinpoint the mm counterparts of these arms \citep{chapillon08, isella10, andrews11, isella13, perez14,marino15}. Also, the parent disks are recognized as transitional disks \citep{espaillat14}, with a giant cavity at the center.

The origin of the observed spiral arms is unclear, though they have been widely speculated to be density waves excited by embedded unseen planets. In linear theory of density waves, the pitch angle, defined as the angle between the azimuthal direction and the tangent of the waves, is set by the aspect ratio $h/r$ of the disk, thereby the disk temperature \citep{rafikov02,muto12}. Fitting the shape of observed arms to linear theory and assuming that the planets are interior to arms has sometimes led to unphysically high disk temperature \citep{benisty15}. Recently, in a pioneering study \citet{juhasz15} calculated the surface density structures of waves using 2D locally isothermal hydro simulations. The resulting disk structures were puffed up in the vertical dimension assuming a Gaussian profile, and synthesized NIR images were produced using 3D radiative transfer simulations. It was concluded that the amplitude of the waves induced by a 1$\mj$ planet at 25~AU in a typical disk is too small to be visible with current NIR facilities.

In this paper following \citet{dong15a}, we revisit the observational appearance of planet-induced density waves in NIR imaging observations. The primary question we address is whether planets can drive density waves that resemble the directly imaged spiral arms. To do this, we carry out 3D global hydrodynamics simulations using the Athena++ code to calculate the density structures of disks. Both locally isothermal (ISO) and adiabatic (ADI) equations of states are explored, and we highlight both the inner and outer density waves. The resulting 3D hydro models are fed into 3D Monte Carlo Radiative Transfer (MCRT) simulations to produce synthetic model images at $1.6\micron$ ($H$~band) using the \citep{whitney13} code. The hydro and MCRT simulations are introduced in Section~\ref{sec:setup}; the modeling results are presented in Section~\ref{sec:results}, followed by a short summary in Section~\ref{sec:summary}.


\section{Simulation Setup}\label{sec:setup}

\subsection{Hydrodynamical Simulations}\label{sec:hydro}

The detailed discussion of the hydro simulations using the Athena++ code and the density wave structure will be presented in Zhu et al. (in prep). Here we provide a short summary of the models. Athena++ is the newly developed grid based code using a higher-order Godunov scheme for MHD and  the constrained transport (CT) to conserve the divergence-free property for magnetic fields (Stone in prep.). Compared with its predecessor Athena \citep{gardiner05,gardiner08,stone08},  Athena++ is highly optimized for efficiency and uses flexible grid structures, allowing global numerical simulations spanning a large range of scales. We have carried out global 3-D simulations in spherical coordinates $r$ (radial), $\theta$ (polar), and $\phi$ (azimuthal) with both isothermal and adiabatic equation of states. In the adiabatic runs the adiabatic index $\gamma$ is chosen to be 1.4.

The initial disk temperature $T$ is constant on cylinders
\begin{equation}
T(r,z)=T_{0}\left(\frac{r}{r_{0}}\right)^{-1/2}\,.
\end{equation}
Therefore the initial disk scale height $h=c_{\rm s}/\Omega\propto r^{1.5}T^{0.5}\propto r^{1.25}$ ($c_{\rm s}$ is the sound speed and $\Omega$ is the angular frequency in the disk), and $h/r=0.1$ at 50~AU. The disk density and azimuthal velocity are set to maintain hydrostatic equilibrium \citep[e.g.][]{nelson13}. The initial disk surface density $\Sigma_0$ decreases as $r^{-1}$ in all our simulations.

The grids are uniformly spaced in log $r$, $\theta$,  $\phi$ with 256$\times$128$\times$688 grid cells in the domain of [15, 150 (AU)]$\times$[$\pi/2$-0.6, $\pi/2$+0.6 ]$\times$[0, 2$\pi$].  Constant $\alpha$ viscosity with $\alpha=10^{-4}$ has been applied. At inner and outer boundaries, all quantities are fixed at the initial states. We have run the simulations for 20 planetary orbits. The planet is on a fixed circular orbit, and there is no accretion from the disk onto the planet. The time scale is chosen so that density waves have established throughout the whole disk (longer than the sound crossing time) while a big and deep gap has not been opened (and consequently no vortex generated at the gap edge). The gap opening process depends on the level of turbulent stress in the disk, which is poorly constrained. In total, we carry out 6 models, and their properties are summarized in Table~\ref{tab:models}.

\subsection{Monte Carlo Radiative Transfer Simulations}\label{sec:mcrt}

Density distributions obtained in our hydro simulations were post-processed via the 3D MCRT calculations using the code developed by \citet{whitney13}, which has been used to model protoplanetary disks in the past (e.g. \citealt{hashimoto12,zhu12,dong12cavity,dong12pds70,follette13,grady13,dong15a}). In MCRT simulations, photons from the central star are absorbed/reemitted or scattered by the dust in the surrounding disk. The temperature in each grid cell is calculated based on the radiative equilibrium algorithm described in \citet{lucy99}. The anisotropic scattering phase function is approximated using the Henyey-Greenstein function \citep{henyey41}. Polarization is calculated assuming a Rayleigh-like phase function for the linear polarization \citep{white79}. Full resolution synthesized PI images at $H$~band ($1.6\micron$) are produced for all models with a pixel size of 0.7~AU/pixel\footnote{In this work, the physical quantity recorded in all model images is the specific intensity in unit of [mJy~arcsec$^{-2}$], or [ergs~s$^{-1}$~cm$^{-2}$~Hz$^{-1}$~arcsec$^{-2}$].}. These images are then convolved by a Gaussian point spread function (PSF) with a full width half maximum (FWHM) of $0.06\arcsec$ and assuming a distance of 140~pc, to achieve an angular resolution comparable with NIR direct imaging observations using Subaru, VLT, and Gemini (the FWHM of a theoretical airy disk is $1.028\lambda/D\sim0.04\arcsec$ for a primary mirror with a diameter $D=8.2$~m at $\lambda=1.6~\micron$).

The MCRT simulation setup is similar to \citet{dong15a}. We construct a 3D disk structure in spherical coordinates with the same grid structure as in the hydro models ($\phi=0$ is west in all images). The central source is a typical Herbig Ae/Be star with a temperature of $10^4$~K and a radius of $2R_\odot$. All simulations are run with 3 billion photon packages. The dust grains in the disk are assumed to be interstellar medium (ISM) grains \citep{kim94} made of silicate, graphite, and amorphous carbon. Their size distribution is a smooth power law in the range of 0.002-0.25~$\micron$ followed by an exponential cut off beyond 0.25~$\micron$. These grains are small enough so that they are dynamically well coupled to the gas, and thus have a volume density linearly proportional to the gas density. The optical properties of the grains can be found in Figure 2 in \citet{dong12cavity}. The total mass of the ISM grains is assumed to be $2\times10^{-5}\msun$\footnote{This corresponds to, for example, a total gas disk mass of 0.02~$\msun$, a 100:1 gas-to-dust-mass-ratio, a $10\%$ dust mass fraction in the small ISM grains, and the rest 90\% in the large grains that have already settled to the disk midplane and do not affect NIR scattering.}.


\section{Modeling Results}\label{sec:results}

In this section we present the modeling results. The artificial central cavity in the hydro models produces a bright rim structure at its edge in scattered light images, which contaminates disk structure within $\sim35$~AU from the center in convolved images. Therefore, we focus only on the outer arms in the $\rp=50$~AU models and the inner arms in the $\rp=125$~AU models, which are all located at $r\gtrsim50$~AU.

\subsection{Density Structure of the Models}\label{sec:results-density}

\begin{figure}
\begin{center}
\includegraphics[trim=0 0 0 0, clip,width=0.99\textwidth]{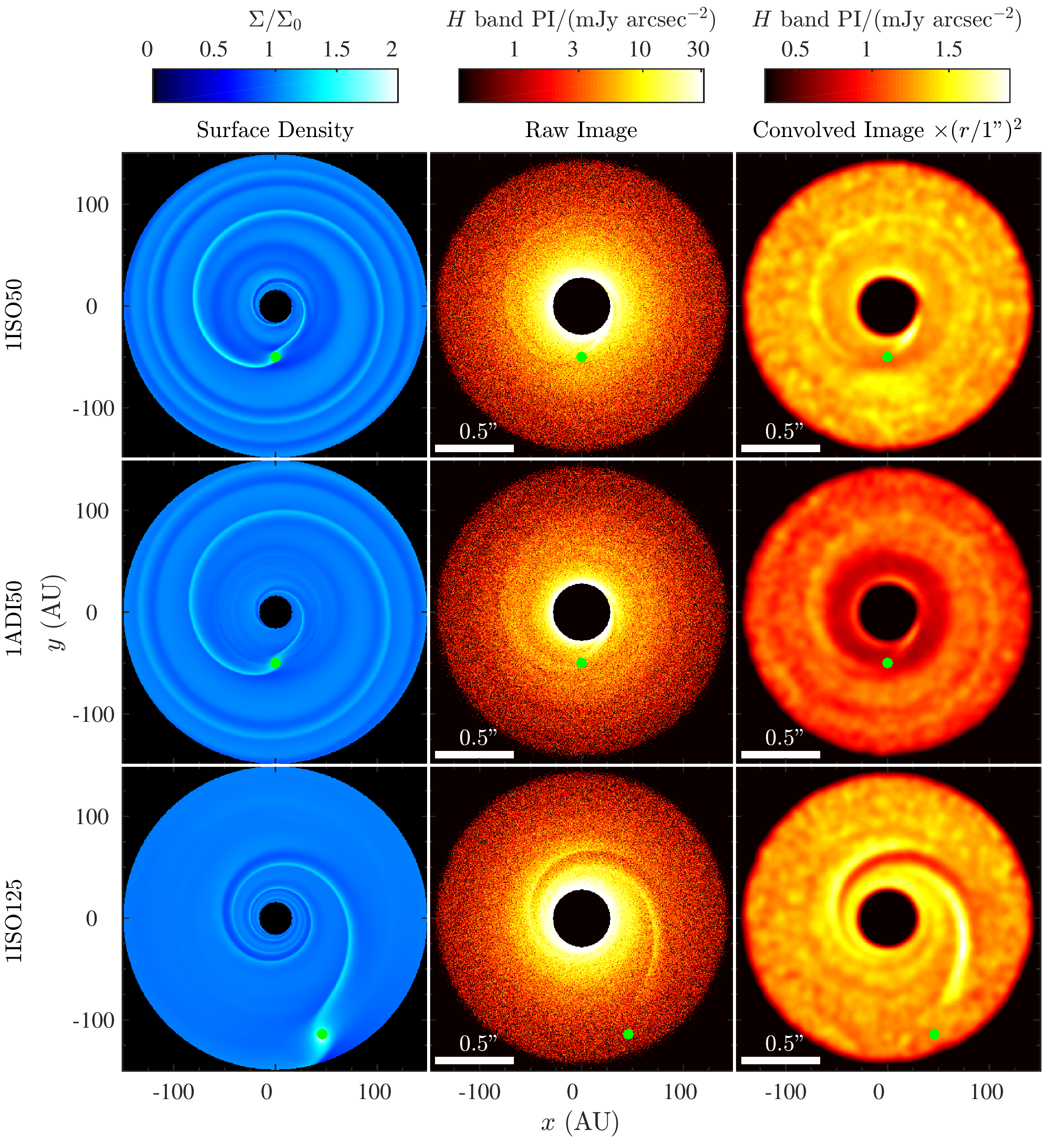}
\end{center}
\figcaption{Model results of the $\mplanet=1\mj$ models, showing the relative surface density perturbation $\Sigma/\Sigma_0$ (left column, linear scale); $H$ band (1.6~$\micron$) full resolution polarized intensity images (middle, logarithmic scale, the central $0.2\arcsec$ has been masked out to highlight the structures in the outer disk); and $1/r^2$-scaled convolved images (right, linear scale, convolved by a Gaussian PSF with FWHM$=0.06\arcsec$). Note that gaps have not been fully opened by the planets. 
\label{fig:1mj}}
\end{figure}

The surface density perturbation $\Sigma/\Sigma_0$ (where $\Sigma_0$ is the initial surface density) of the hydro models are shown in the left column in Figures~\ref{fig:1mj} and \ref{fig:6mj}. When $\mplanet/M_\star\gtrsim(h/r)^3$, a planet excites one inner and one outer waves originated from the planet (the primary arms from now on) and another two shifted by roughly $180^\circ$ in the azimuthal direction (the secondary arms from now on)\footnote{The two spiral arms are seen in previous simulations, e.g. Figure 10 of \citet{devalborro06}, and commented in \citet{juhasz15}.}, while their strengths depend on $\mplanet$, $h/r$, and EOS. The primary waves are always stronger than the secondary waves, which are practically not recognizable in 1ISO50 and 1ADI50 runs but visible in all other cases. In $1\mj$ models, $\Sigma_{\rm wave}$ never exceeds $\Sigma_0$ by $50\%$, consistent with \citet{juhasz15}, while in the $6\mj$ models the surface density enhancement of the primary waves can be more than $100\%$ (Table~\ref{tab:models}). $\Sigma$ enhancement in ISO models is close to twice that in the corresponding ADI models. We note that since in 3D the disk is not hydrostatic in the vicinity of the shock, its vertical structure cannot be fully described by puffing up the surface density distribution from 2D simulations assuming hydrostatic (Gaussian) vertical density profile, as performed in \citet{juhasz15}.

\subsection{Scattered Light Images of the Outer Arms}\label{sec:results-outerarms}

\begin{figure}
\begin{center}
\includegraphics[trim=0 0 0 0, clip,width=0.99\textwidth]{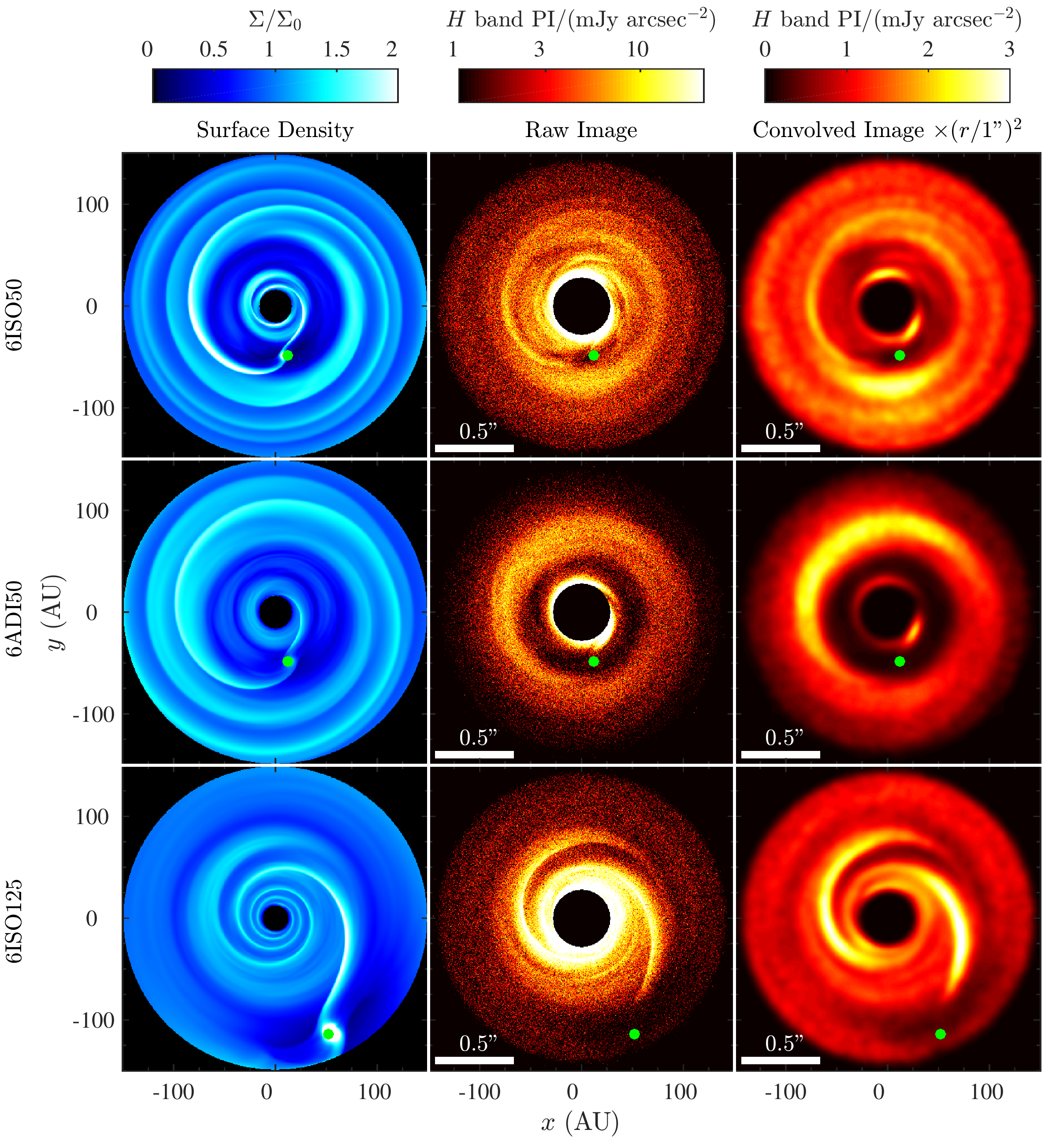}
\end{center}
\figcaption{The same as Figure~\ref{fig:1mj}, but for the 6~$\mj$ models. 
\label{fig:6mj}}
\end{figure}
\begin{figure}
\begin{center}
\includegraphics[trim=0 0 0 0, clip,width=0.7\textwidth]{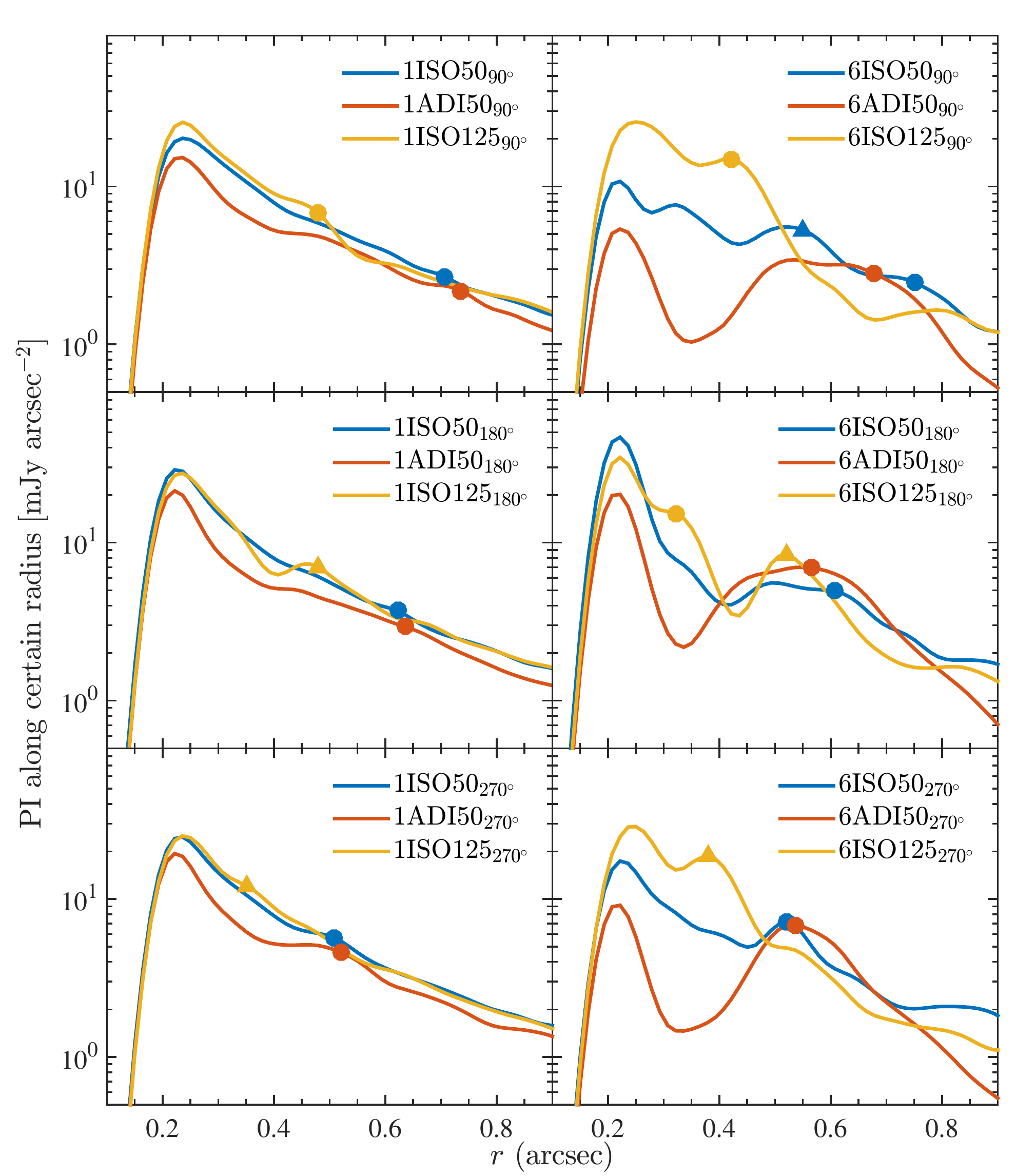}
\includegraphics[trim=0 0 0 0, clip,width=0.29\textwidth]{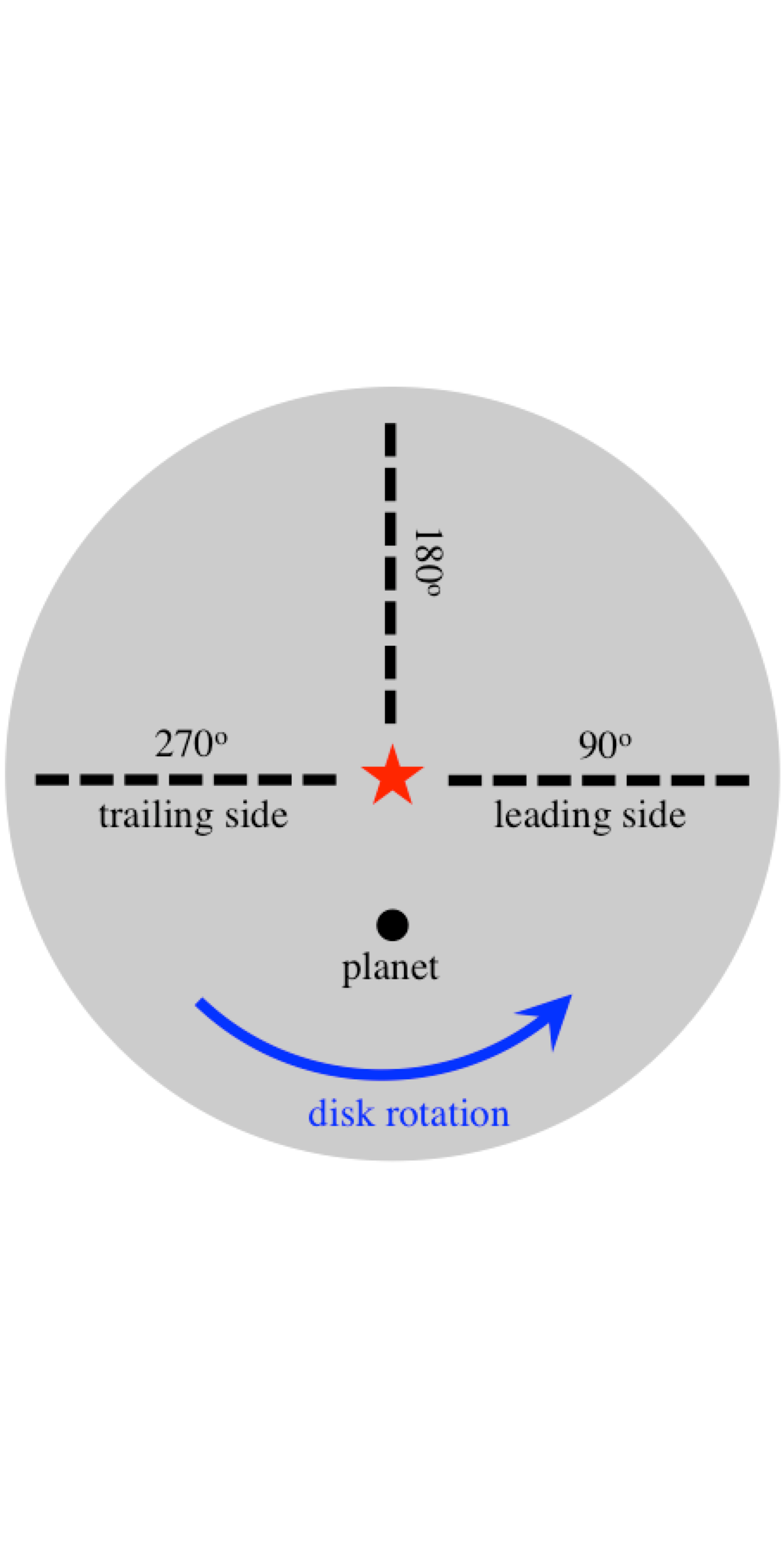}
\includegraphics[trim=0 0 0 0, clip,width=0.7\textwidth]{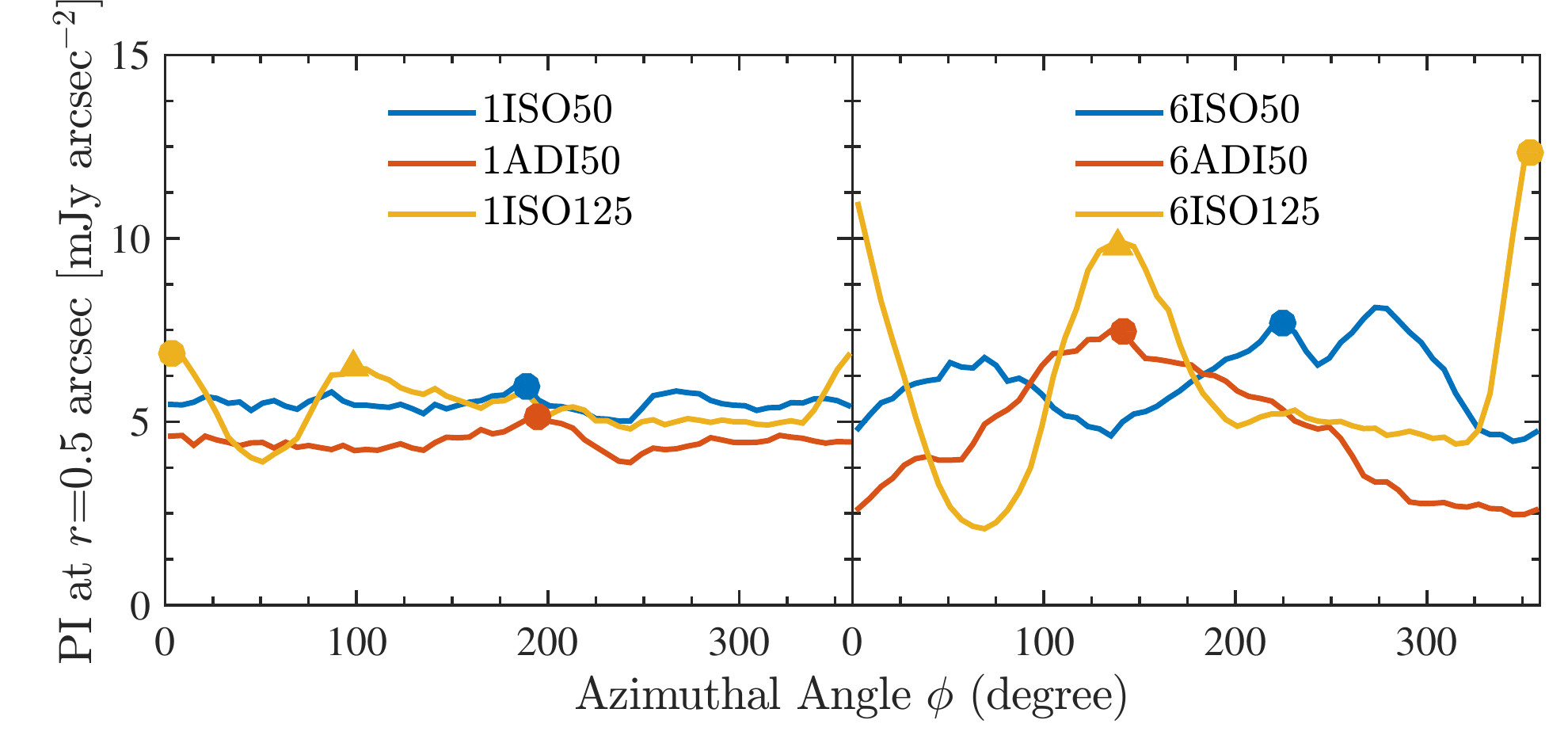}
\includegraphics[trim=0 0 0 0, clip,width=0.29\textwidth]{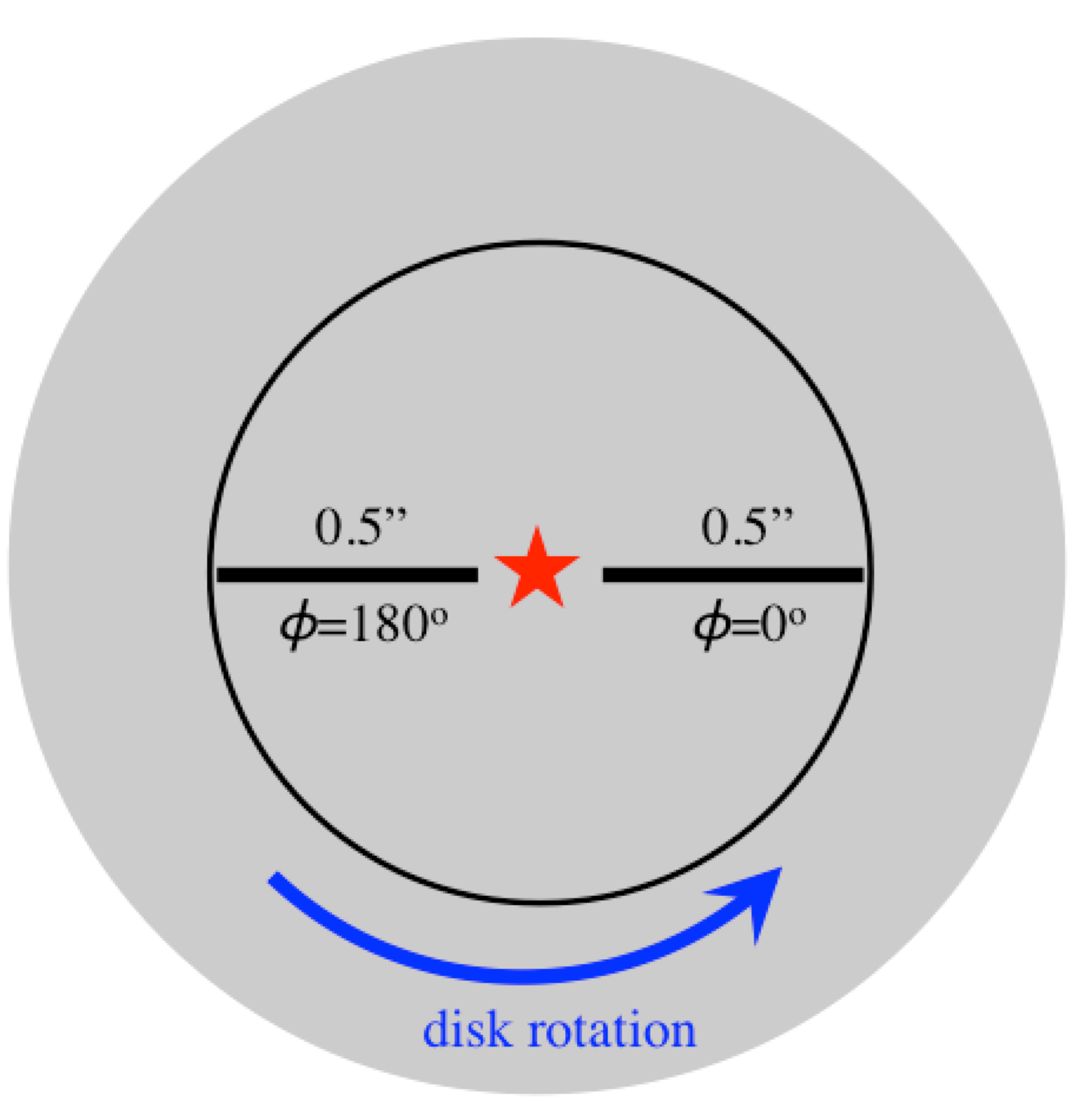}
\end{center}
\figcaption{Top: Radial profiles of convolved images at $90^\circ$, $180^\circ$, and $270^\circ$ away from the planet, as illustrated in the cartoon. Bottom: Azimuthal profiles of convolved images at $r=0.5\arcsec$ (70~AU) from the star. The (dots, triangles) mark the intersections with the (primary, secondary) arms. 
\label{fig:image_profiles}}
\end{figure}

Full resolution and convolved $H$~band images for all models are shown in Figure~\ref{fig:1mj} and \ref{fig:6mj}. The outer arms in the $1\mj$ models are practically not traceable in both full resolution images and convolved images, echoing \citet{juhasz15} (they are recognizable in the $1/r^2$-scaled convolved images as the dynamical range of the background is largely suppressed). The peak surface brightness enhancement along the waves is $\sim10-20\%$ (Table~\ref{tab:models}), producing only marginal fluctuations on radial and azimuthal image profiles (Figure~\ref{fig:image_profiles}). The $6\mj$ models produce much more prominent arms (Figure~\ref{fig:6mj} and \ref{fig:image_profiles}). The primary arm in 6ADI50 can be up to $\sim$2 times brighter than the background, and it is radially extended, with a radial width $\sim0.2\arcsec$ at $r=0.5\arcsec$. Note that the contrast of arms in our convolved images are expected to be higher than in observations due to (1) model images are convolved by a Gaussian PSF, which has a similar kernel but less extended wings than a realistic PSF, and (2) no observational noises and instrument effects are included, which will further weaken the contrast of features. The 2 outer arms have a rough $m=2$ symmetry, and are tightly wound, with pitch angles no bigger than $6^\circ$ (Table~\ref{tab:models}), which is roughly half the values in observed systems (listed in Table~\ref{tab:models} as well for comparison). They may not be distinguishable from a ring structure in observations with low S/N.

An interesting finding is that arms in ADI models are more prominent than in ISO models. The arm surface brightness enhancement in 1ISO50 is only half of that in 1ADI50, despite having a surface density enhancement 2 times larger. $6\mj$ models show similar effects. This is caused by a combination of shock heating and hydraulic effects in waves. As predicted by \citet{goodman01,rafikov02} and shown in simulations \citep{dong11b,duffell13,zhu13,richert15}, density waves shock at a couple of scale heights away from the planet's orbit when $\mplanet\gtrsim M_\star(h/r)^3$. Shock heating heats material along the waves in ADI cases, resulting in a higher scale height than in ISO cases and puffing up the waves in the vertical direction ($h=c_{\rm s}/\Omega\propto\sqrt{T}$). The hydraulic effects seen in 3D simulations of wave shocks by \citet{boley06} also makes ADI waves more puffed up. Also interestingly, in the $\mplanet=6\mj$ models while both arms are visible in 6ISO50 with roughly equal strengths (also note the brightening behind the planet, \citealt{jang-condell12}), in 6ADI50 the primary arm is much more prominent. This may indicate a difference in the excitation and dynamics between the two arms, as the shock heating and hydraulic effects are more effective in the primary arm. 

Lastly, we perform experiments in which we collapse our 3D hydro disks to 2D, then puff them up back to 3D assuming Gaussian profiles in the vertical structure, similar to what was done in \citet{juhasz15}. Results show that a full 3D hydro treatment does not affect the detectability in the outer arms excited by a $1\mj$ planet, however in $\mplanet=6\mj$ models it does produce noticeable differences in images, particularly for the inner arms. A more detailed discussion can be found in Zhu et al. (in prep).

\subsection{Scattered Light Images of the Inner Arms}\label{sec:results-innerarms}

In our simulations the inner arms are open and prominent. In convolved images, the arms are up to $30-40\%$ brighter than the background region at the same radius in 1ISO125, and $150-160\%$ brighter in 6ISO125 (Table~\ref{tab:models}). Interestingly, although the secondary arm are much weaker than the primary in surface density contrast, the two are roughly equally bright in scattered light images. In the azimuthal direction, the primary and secondary arms extends over $\sim180^\circ$ and $\sim270^\circ$, respectively. The 2 arms are roughly $180^\circ$ rotationally symmetric (in both location and contrast). The pitch angle is about $15^\circ$ for the primary arm and $10^\circ$ for the secondary. The typical FWHM of the arms in the radial direction is $\sim0.1\arcsec$. 

\begin{figure}
\begin{center}
\includegraphics[trim=0 0 0 0, clip,width=0.45\textwidth]{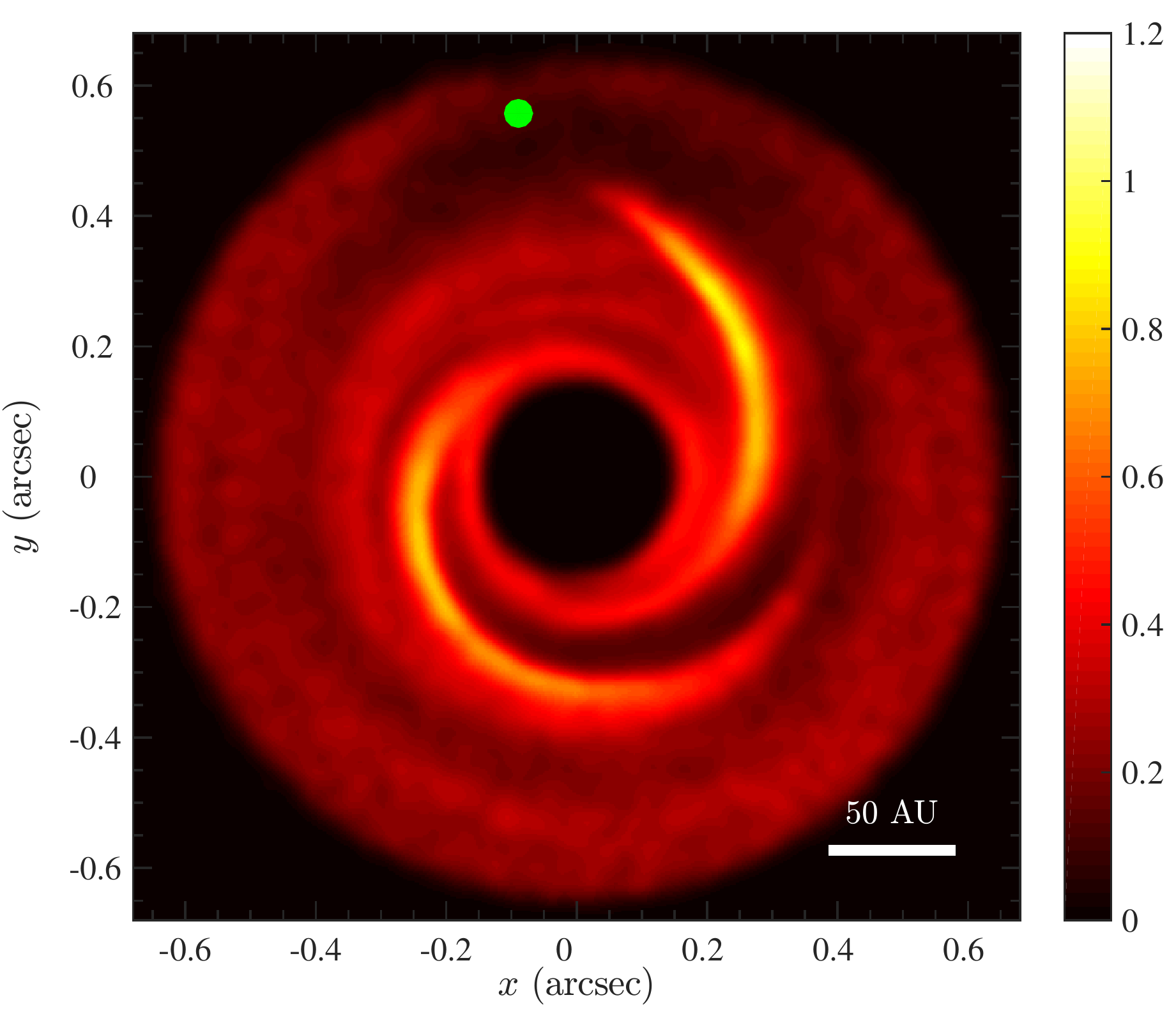}
\includegraphics[trim=0 0 0 0, clip,width=0.54\textwidth]{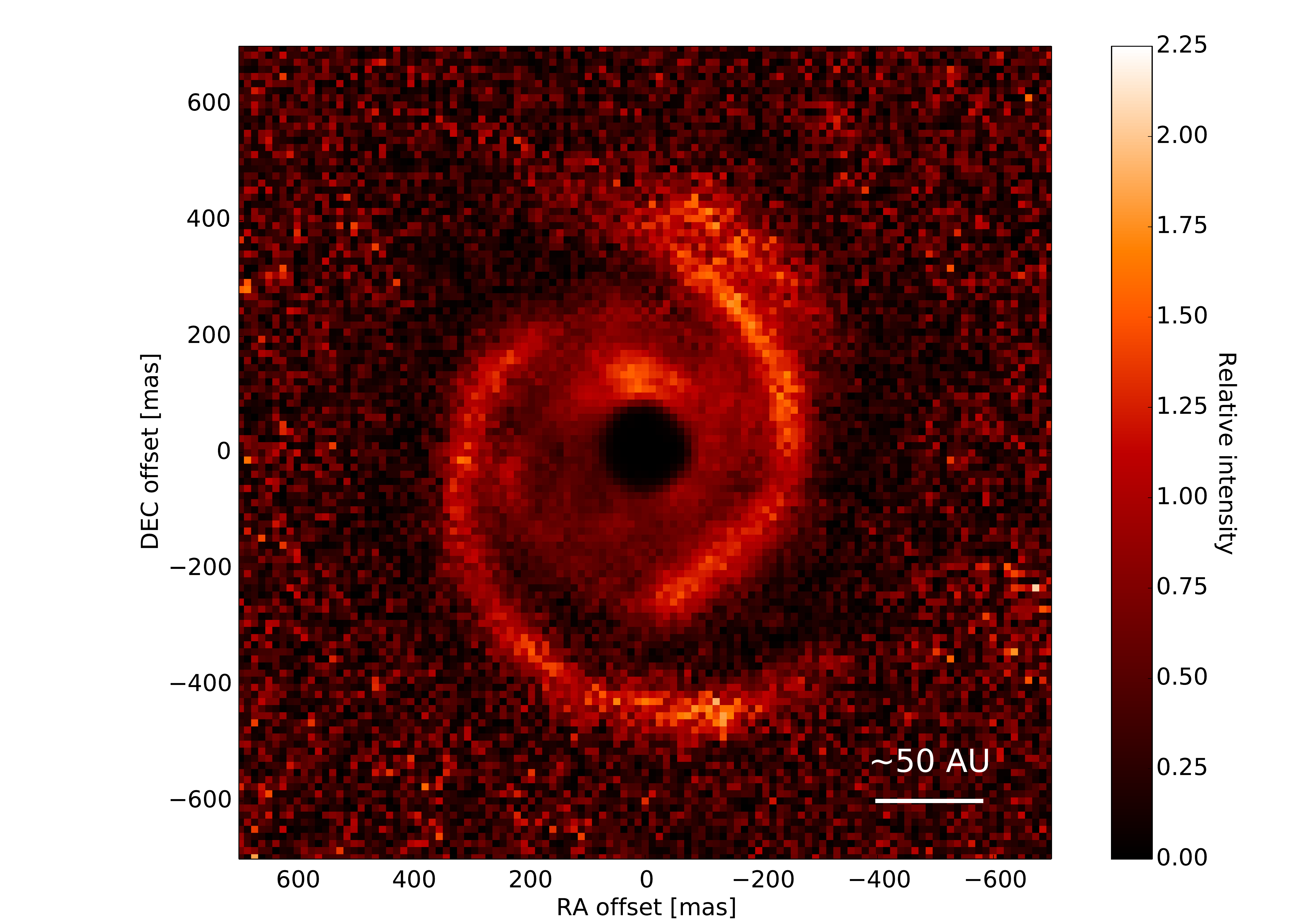}
\end{center}
\figcaption{Comparison between 6ISO125 (left, $1/r^2$-scaled convolved image) and MWC~758 (right, \citealt{benisty15}). The model is rescaled to $\rp=158$~AU, and the Gaussian PSF convolution adopts the source distance (280~pc) and the angular resolution of the observation (FWHM=$0.03\arcsec$). Units are arbitrary. The green dot in the model image marks the location of the planet ($\mplanet/M_\star=6\times10^{-3}$).
\label{fig:comparison}}
\end{figure}

In general, the morphology of the inner arms is quite similar to the observed ones in SAO~206462 \citep{muto12,garufi13} and MWC~758 \citep{grady13,benisty15}, in terms of (1) the rough $m=2$ symmetry,  (2) pitch angle, and (3) azimuthal extension. We note that the shape (i.e. pitch angle) of these inner arms cannot be described by the linear theory in \citet{rafikov02,muto12}, as the waves are highly nonlinear. A comparison between MWC758 and a rescaled 6ISO125 model with the actual source distance and angular resolution is shown in Figure~\ref{fig:comparison}. In addition, for observed arms while it is difficult to measure the surface brightness enhancement (difficulty in defining a true ``background'') and radial width (noises in the data) in an equally clean way as we do to our theoretical models here, observed arms are mostly in the bulk part $100-300\%$ brighter than the background and around $0.1\arcsec$ in width in the radial direction, broadly consistent with the 6ISO125 model. In general, the morphology of the inner arms in the 6ISO125 run is tantalizingly similar to the observed ones.



\section{Summary}\label{sec:summary}

Our main conclusions are:

\begin{enumerate}

\item The inner spiral arms excited by a massive planet ($q=\mplanet/M_\star=6\times10^{-3}$) can be visible in scattered light images under current NIR direct imaging capability, and they have a morphology resembling the observed spirals, such as in the SAO~206462 \citep{muto12,garufi13} and MWC~758 \citep{grady13,benisty15} systems (Figure~\ref{fig:comparison}). They exhibit (1) roughly $m=2$ symmetry, (2) pitch angles in between $10-16^\circ$, (3) an azimuthal extent of $\sim180-270^\circ$, and (4) $\sim150\%$ surface brightness enhancement relative to the background, all broadly consistent with observations. Note that the shape of these arms cannot be fully described by the weakly non-linear theory \citep{rafikov02}.

\item The outer spiral arm excited by a $q=6\times10^{-3}$ planet can also be visible. However they cannot explain the observed spiral arms, as they are too tightly wound given typical disk scale height (pitch angle $\lesssim4^\circ$, 2.5 times or more smaller than observations), which has been noted in the case of MWC~758 by \citet{benisty15}. In addition, the contrast of the isothermal waves is lower than observed spiral arms, while the more prominent ADI waves are too extended in the radial direction, with a radial width a factor of 2 or more larger than observations.

\item Planet induced density waves have a higher surface density enhancement in isothermal disks than in adiabatic disks. However, in scattered light images the arms in adiabatic disks are more prominent, thanks to a more extended structure in the vertical direction due to shock heating and possibly hydraulic effects. Also, 3D effects in isothermal waves can be important (Zhu et al. in perp.)

\end{enumerate}

One key factor in the appearance of planet-induced density waves may be that the gap cannot be fully opened by the planet given the duration of our simulations, so that the bulk part of waves are not located in a deep gap region. In our models, the planet is located roughly at a distance 1.5-3 times the main part of the inner arms. If observed arms are driven by a companion on a circular orbit outside the arms, then a $\sim10\mj$ companion (possibly a brown dwarf) is needed at around 100 AU in SAO~206462 (at a distance of 140~pc, \citealt{muto12}), and 160~AU in MWC~758 (at a distance of 280~pc, \citealt{grady13}). They can be excellent targets for direct imaging observations.. However, the waves may also be excited by a recent flyby, or a companion on a very eccentric orbit. Lastly, it has been speculated that the cavities in transitional disks are opened by multiple giant planets. Observed global-scale $m=2$ arms so far are all found outside the central cavities in transitional disks. The hypothetical outer companion suggested in our work cannot be associated with the putative cavity-opening planets; the latter would reside inside the cavity and their density waves are unlikely to be detectable.


\section*{Acknowledgments}

We thank Eugene Chiang, Barbara Whitney, Eric Pantin, Eduard Vorobyov, and Jun Hashimoto for insightful discussions and help in this work. We also thank the anonymous referee for constructive suggestions that largely improved the quality of the paper. We thank Myriam Benisty for kindly sharing with us the VLT/SPHERE image of MWC~758. This project is partially motivated by the Subaru based SEEDS program (PI: M. Tamura). This project is supported by NASA through Hubble Fellowship grants HST-HF-51333.01-A (Z.Z.) and HST-HF-51320.01-A (R.D.) awarded by the Space Telescope Science Institute, which is operated by the Association of Universities for Research in Astronomy, Inc., for NASA, under contract NAS 5-26555. All hydrodynamic simulations are carried out at the Texas Advanced Computing Center (TACC) at The University of Texas at Austin using Stampede through XSEDE grant TG-AST130002.




\begin{landscape}
\begin{table}
\footnotesize
\begin{center}
\caption{Model Properties}
\begin{tabular}{cccccccc}
\tableline\tableline
Model/Observation  & EOS & $q$ & $\rp$ & $ (\Sigma_{\rm wave,peak}-\Sigma_0)/\Sigma_0$ & $(I_{\rm wave,peak}-I_{\rm b})/I_{\rm b}$  &  Radial FWHM & Pitch Angle \\
& & $\times10^{-3}$ & AU & $\%$ & $\%$ & arcsec & degree \\
(1) & (2) & (3) & (4) & (5) & (6) & (7) & (8) \\
\tableline
1ISO50 & ISO & 1 & 50 & 49(OP) & 10(OP) & N/A & 6(OP) \\
1ADI50 & ADI & 1 & 50 & 28(OP) & 20(OP) & N/A & 6(OP) \\
1ISO125 & ISO & 1 & 125 & 31(IP), 21(IS) & 40(IP), 30(IS) & $\sim0.09$(IP), $\sim0.08$(IS) & 16(IP),10(IS) \\
6ISO50 & ISO & 6 & 50 & 150(OP),61(OS) & N/A & $\sim0.09$(OP), $\sim0.1$(OS) & 4(OP),2(OS) \\
6ADI50 & ADI & 6 & 50 & 77(OP), 36(OS) & 190(OP) & $\sim0.24$(OP) & 4(OP) \\
6ISO125 & ISO & 6 & 125 & 125(IP), 48(IS) & 160(IP), 150(IS) & $\sim0.1$(IP), $\sim0.08$(IS) &15(IP),10(IS) \\
\tableline
SAO~206462\tablenotemark{a} &&&&& $\sim100$ (east), $\sim300$ (west) & $\sim0.11$(east), $\sim0.13$(west) & 11(east), 10(west)\\
MWC~758\tablenotemark{a} &&&&& $\sim200$ (east), $\sim250$ (west) & $\sim0.07$(east), $\sim0.1$(west) & 11(east), 10(west)\\
\tableline
\end{tabular}
\tablecomments{Properties of the models. (1) Model name. (2) Equation of state in the hydro simulations: isothermal (ISO) or adiabatic (ADI). (3) $q=\mplanet/M_\star$. For systems around a $1M_\odot$ star the list number is just planet mass in $\mj$. (4) We only look at outer arms in the $\rp=50$~AU models and inner arms in the $\rp=125$~AU models. (5) Peak surface density enhancement on the waves. The letters in parenthesis are for: (O) -- outer arms; (I) -- inner arms; (P) primary arms (originated from the planet); (S) -- secondary arms. The secondary arm in 1ISO50 and 1ADI50 are not recognizable. (6) Peak surface brightness enhancement on the waves in convolved images. $I_{\rm b}$ is the surface brightness in the background region around position angle $\phi=270-325^\circ$ at the same distance from the center as the arms (we avoid the dark shadowed regions right behind the arms). The two outer arms in 6ISO50 are both prominent and tightly winded up, which makes it difficult to define a ``background'' at the distance of the arms. The secondary arm in 6ADI50 cannot be easily distinguished from the background. (7) FWHM of the main part of the arms in the radial direction. (8) Averaged pitch angle of the arms. (a) The more recent and sharper VLT images in \citet[SAO~206462]{garufi13} and \citet[MWC~758]{benisty15} are used for the pitch angle and radial width measurements, while older Subaru datasets in \citet[SAO~206462]{muto12} and \citet[MWC~758]{grady13} are used for surface brightness enhancement measurements as the absolute polarized intensity calibration is not always available in the new VLT datasets \citep{benisty15}. We only focus on the 2 major arms on the east and west sides in each case. It is sometimes difficult to perform measurements on observed arms in an equally clean way as we do to our models due to observational noises and time variations.}
\label{tab:models}
\end{center}
\end{table}
\end{landscape}


\end{document}